# Crosstalk-free Conjugate Networks for Optical Multicast Switching

Yun Deng, *Student Member, IEEE,* Tony T. LEE, *Fellow, IEEE*

*Abstract*— High-speed photonic switching networks can switch optical signals at the rate of several terabits per second. However, they suffer from an intrinsic crosstalk problem when two optical signals cross at the same switch element. To avoid crosstalk, active connections must be node-disjoint in the switching network. In this paper, we propose a sequence of decomposition and merge operations, called conjugate transformation, performed on each switch element to tackle this problem. The network resulting from this transformation is called conjugate network. By using the numbering-schemes of networks, we prove that if the route assignments in the original network are link-disjoint, their corresponding ones in the conjugate network would be node-disjoint. Thus, traditional nonblocking switching networks can be transformed into crosstalk-free optical switches in a routine manner. Furthermore, we show that crosstalk-free multicast switches can also be obtained from existing nonblocking multicast switches via the same conjugate transformation.

## I. INTRODUCTION

OWING to the explosive growth of the Internet traffic, there are increasing demands for the transmission capacity and faster switching technologies in telecommunication networks. The development of optical devices and the deployment of all-optical networks (AONs) have drawn more and more attentions. The optical switching networks we are focusing on serve a heterogeneous population of users who require both guaranteed bandwidth (GBW) connections and bandwidth on demand (BOD) services of differing average information rates and burstiness. To serve these users, the switch provides point-to-point and point-to-multipoint services between the access stations. In the face of optical technology evolution, the switch architecture should seamlessly support the addition of even higher speed stations in the future.

A $2 \times 2$ switch node may use a directional coupler (DC) whose coupling ratio is changed by varying the refractive index of the material in the coupling region. One commonly used material is lithium niobate (LiNbO$_3$). Such an electrooptic switch is capable of changing its state extremely rapidly, typically in less than a nanosecond.

Therefore, high speed optical switching networks can be constructed by using those DC devices as basic building blocks. The major obstacle, however, associated with low cost

Manuscript received March 6, 2006; revised June 26, 2006. This work was supported by the Research Grant Council of Hong Kong under Earmark Grants CUHK 414305 and CUHK 4380/02E.

Yun Deng and Tony T. Lee are with the Department of Information Engineering , the Chinese University of Hong Kong, Shatin, HKSAR, China (email:ydeng1@ie.cuhk.edu.hk; ttlee@ie.cuhk.edu.hk)

Digital Object Identifier 00.0000/JLT.2006.00000

DCs is the crosstalk problem [1], in which a portion of optical power of one signal may be coupled into another signal when those two optical signals pass through the same DC node simultaneously. To avoid crosstalk problem completely, all I/O paths in the switching network must be node-disjoint, which is different from the link-disjoint requirement in the traditional nonblocking switching networks.

Vertical stacking [2] of multistage interconnection networks (MINs) [3] is a straightforward technique for creating the crosstalk-free optical switching networks, in which multiple parallel switching planes are provided and connections are assigned to different planes to avoid potential crosstalk. However, the number of planes needed for crosstalk-free routing is $O(\sqrt{N})$ which is too large to be practical, where $N$ is the number of input/output ports. It is shown in [4], [5] that relaxing the crosstalk constraint or increasing the length of each switching plane can reduce the complexity. The strictly crosstalk-free conditions under different constraints are given in [4]. The minimal number of planes needed are derived in [5] for given number of stages in a MIN. Rearrangeably crosstalk-free conditions are discussed in [6], [7]. The widesense nonblocking networks under a crosstalk constraint is introduced in [8]. The parallel routing algorithm of strictly or rearrangeably design is discussed in [9], [10].

By changing the vertically stacking from space domain to time domain, a crosstalk-free scheduling algorithm is described in [11]–[13]. A wavelength approach is proposed in [14], [15], in which the crosstalk between signals carried in different wavelengths can be filtered at the destination.

The performance of vertical stacking Banyan networks under blocking situation is evaluated in [16]–[19], in which [16] demonstrated the simulation results, the upper and lower bounds of blocking probability with respect to the number of switch planes are derived in [17], and the lower bound with respect to the length of each switching plane is given in [18], while [19] proposed an analytical model under random routing strategy to manifest the tradeoff between the blocking probability and hardware cost.

A bipartite graph representation of crosstalk-free route assignments in MINs is discussed in [20], [21], in which the links are represented by vertices and switching elements are represented by edges. This representation demonstrates the correspondence between crosstalk-free route assignments and nonblocking route assignments. However, algorithms and proofs of legitimate route transformation between the MIN and its bipartite graph representation are not available.

[22] studied the parallel Banyan networks, respectively, nonblocking in the strict-sense and the wide-sense for the

multicast connections. A class of wide-sense nonblocking multicasting networks is proposed in [23]. To the best of our knowledge, the only result on crosstalk-free multicasting is presented in [24], which is a time domain approach.

Our goal in this paper is to provide an easy-to-implement transformation from traditional nonblocking networks to crosstalk-free optical switching networks. In principle, to completely avoid crosstalk between two optical signals crossing the same DC element, active I/O paths must be node-disjoint in the switching network. Topologically, this problem is similar to the nonblocking route assignments of conventional electronic switching networks, in which active I/O paths must be link-disjoint. We propose a class of networks, called *conjugate networks*, that can be obtained by performing a sequence of decomposition and merge operations on each switch element of an existing network. We show that any nonblocking route assignments in the original network will become node-disjoint in the resulting conjugate network. Therefore, nonblocking route assignments and rearrangements algorithms developed for conventional electronic switching networks can be adopted via the conjugate transformation to implement crosstalk-free optical switches. The complexity of the conjugate network is $d$ times that of the original network constructed by $d \times d$ switching elements.

Furthermore, we show that crosstalk-free multicast switches can also be constructed in a similar manner by applying the conjugate transformation to the generalized multicast switch and the associated nonblocking rank-based route assignment algorithm proposed in [25], [26]. Specifically, this multicast switch architecture can provide rearrangeably crosstalk-free route assignments under the conjugate transformation by taking up the original nonblocking route assignments.

The sequel of this paper is organized as follows. In section II, we introduce the basic concepts of conjugate transformation and conjugate networks. In section III, we employ a numbering-scheme of Benes networks to prove the intrinsic crosstalk-free properties of conjugate Benes networks. We extend the conjugate transformation to three-stage Clos network in section IV and to the crosstalk-free multicast switch in section V, which is synthesized by cascaded combination of a conjugate Benes copy network and a point-to-point conjugate Benes network. Finally, conclusions are summarized in Section VI.

## II. BASIC CONCEPTS OF CONJUGATE TRANSFORMATION AND CONJUGATE NETWORKS

The fundamental concepts of conjugate transformation and conjugate networks are introduced in this section. A $4 \times 4$ Benes network [27] with nonblocking connections, A, B, C and D, are depicted in Fig. 1. We label the upper output link of each node by 0 while the lower one by 1, and define a connection's *link sequence* as the sequence of outgoing link labels along the path of the connection. For example, the link sequence of connection A in Fig. 1 is 011.

The two operations, decomposition and merge, of conjugate transformation are delineated as follows.

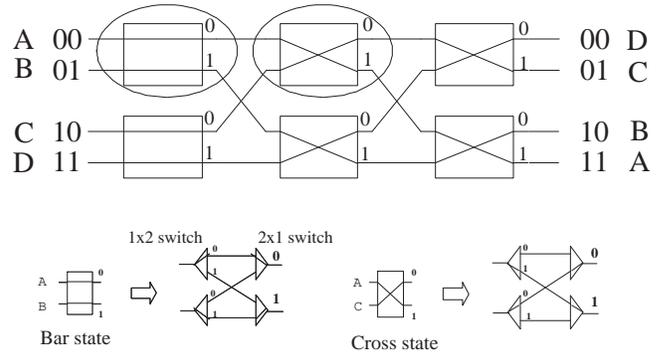

Fig. 1. Conjugate decomposition of Benes network

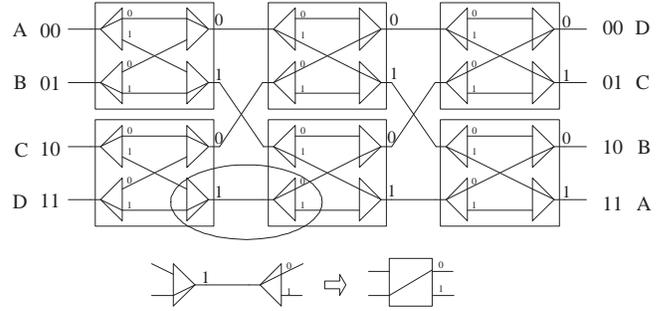

Fig. 2. Conjugate merge of decomposed Benes network

*1) Step 1– decomposition:* In the first step, each node of the network is decomposed into a 2-stage network composed of two $1 \times 2$ and two $2 \times 1$ switch elements as shown in Fig. 1. The routing decision now is made at each of those $1 \times 2$ elements, whose upper and lower output links are labelled respectively by 0 and 1 as usual.

*2) Step 2– merge:* The two adjacent nodes connected by a single link in the decomposed network do not carry any switching functions, they can be merged into a $2 \times 2$ switch node as shown in Fig. 2. It is shown in Fig. 3 that the crosstalk will never occur in a merged node because it only carries at most one signal at any time.

The network resulting from this two-step conjugate transformation is called *conjugate network*. This transformation actually converts each internal link of the original network to a $2 \times 2$ node in the conjugate network. The number of nodes in the conjugate network is roughly two times that of the original network, in which one node is decomposed into four in the first step and two nodes are merged into one, excluding those nodes in the first and last stages, in the second step. In principle, this conjugate transformation applies to any interconnection networks and also to multicast switching.

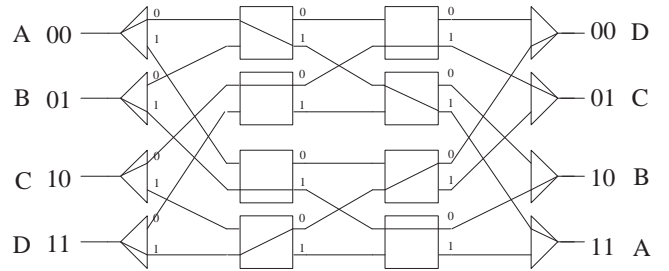

Fig. 3. Conjugate Benes network

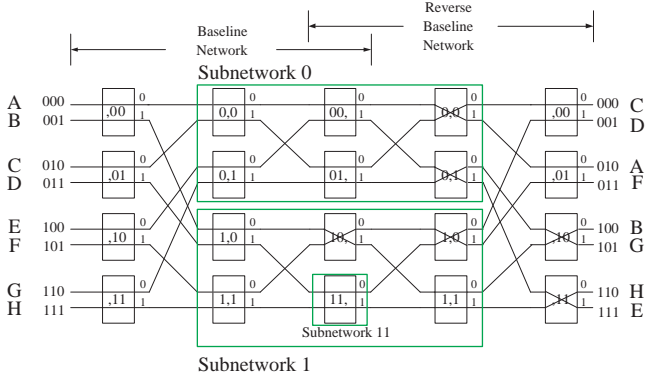

Fig. 4. Numbering of Benes network

The sequence of link labels along a connection path remains the same in the course of decomposition or merge operations. A one-to-one correspondence between paths in an interconnection network and its conjugate network can be established by the numbering scheme of networks. An example of this correspondence is elaborated in the next section to show that the link-disjoint route assignments in the Benes network will become node-disjoint paths in the conjugate network.

## III. BENES NETWORKS AND CONJUGATE BENES NETWORKS

In this section, we first introduce the numbering scheme of interconnection networks, and then provide the algebraic proof of crosstalk-free properties associated with the conjugate Benes network.

### A. Benes networks

The Benes network is constructed recursively according to the three-stage Clos network with 2 central modules. A $N \times N$ Benes network has $2n-1$ stages with $N/2$ modules in each stage, where $n = log_2 N$. An $8 \times 8$ Benes network is shown in Fig. 4, in which the first- and last-stage modules are connected to the subnetworks 0 and 1 that can be further decomposed into $4 \times 4$ three-stage Clos networks with central modules 00 and 01, or 10 and 11, respectively.

It is well-known that the Benes network is rearrangeable nonblocking [27], and the route assignment problem has been extensively studied, such as the sequential looping algorithm [28] with time complexity on the order of $O(N \log N)$, the parallel algorithm of order $O(\log^2 N)$ for full permutations [29] and the one dealing with partial permutations [30].

The numbering scheme of the Benes network is closely related to its recursive construction and defined as follows. The upper and lower output links of each node are labelled by 0 and 1, respectively. As shown in Fig. 4, each node is labelled by a 2-tuple of $(n-1)$ bits total from top to bottom, in which the first binary number is the top-down numbering of the subnetwork and the second one is the top-down numbering of the node within the subnetwork. For example, the node $(0,1)$ in the second stage is the node 1 within the subnetwork 0. The subnetwork numbering part of all nodes in the first and the last stage is empty, denoted by $\phi$, because they are not within in any proper subnetworks. In contrast, the node numbering part of each central module in the middle stage is empty $\phi$, because it is the smallest subnetwork that contains only a single node. In all figures shown in this paper, the empty numbering $\phi$ is omitted without causing any confusions.

Due to the symmetric structure of Benes network, there are two nodes labelled by $(a_1 \ldots a_{i-1}, b_1 \ldots b_{n-i})$, one in stage $i$ and the other one in stage $2n-i$, where $a_1 \ldots a_{i-1}$ is the numbering of the subnetwork and $b_1 \ldots b_{n-i}$ indicates the node numbering within this subnetwork. However, the node $N_{i+1}(a_1 \ldots a_{i-1} a_i, b_1 \ldots b_{n-i-1})$ in stage $i+1$ for $i = 1, \ldots, n-1$ is attached to the node $N_i(a_1 \ldots a_{i-1}, b_1 \ldots b_{n-i})$ in stage $i$ by the output link $a_i$, while the node $N_{2n-i}(a_1 \ldots a_{i-1}, b_1 \ldots b_{n-i-1} b_{n-i})$ in stage $2n-i$ for $i = 1, \ldots, n-1$ is attached to the node $N_{2n-i-1}(a_1 \ldots a_i, b_1 \ldots b_{n-i-1})$ in stage $2n-i-1$ by the output link $b_{n-i}$.

As indicated in Fig. 4, the Benes network is actually formed by two subnetworks, from the input stage to the middle stage is a baseline network followed by its mirror image, a reverse baseline network from the middle stage to the output stage. Both the baseline and reverse baseline networks have the unique-path and self-routing properties. Thus, any path from an input to an output is completely determined by the central module selected, and there are $N/2$ paths, corresponding to $N/2$ central modules, from any input to any output in a Benes network. Based on the node and link numbering scheme, the path connecting the source $S(s_1 \ldots s_n)$, the destination $D(d_1 \ldots d_n)$, and the selected central module $N_n(x_1 \ldots x_{n-1}, \phi)$ can be symbolically expressed as follows:

i) The sub-path within the baseline network:

$$S(s_1 \ldots s_n) \longrightarrow \underbrace{N_1(\phi, s_1 \ldots s_{n-1})}_{\text{stage 1}} \xrightarrow{x_1} \ldots \xrightarrow{x_{i-1}}$$

$$\underbrace{N_i(x_1 \ldots x_{i-1}, s_1 \ldots s_{n-i})}_{\text{stage } i} \xrightarrow{x_i} \ldots \xrightarrow{x_{n-1}}$$

$$\underbrace{N_n(x_1 \ldots x_{n-1}, \phi)}_{\text{stage } n};$$

ii) the sub-path within the reverse baseline network:

$$\underbrace{N_n(x_1 \ldots x_{n-1}, \phi)}_{\text{stage } n} \xrightarrow{d_1} \ldots \xrightarrow{d_{n-i}}$$

$$\underbrace{N_{2n-i}(x_1 \ldots x_{i-1}, d_1 \ldots d_{n-i})}_{\text{stage } 2n-i} \xrightarrow{d_{n-i+1}} \ldots \xrightarrow{d_{n-1}}$$

$$\underbrace{N_{2n-1}(\phi, d_1 \ldots d_{n-1})}_{\text{stage } 2n-1} \xrightarrow{d_n} D(d_1 \ldots d_n),$$

The link sequence of this connection is $x_1 \ldots x_{n-1} d_1 \ldots d_n$. For example, the path of connection B, shown in Fig. 4, from input 001 to output 100 through the central module $(10, \phi)$ is given as follows:

$$S(001) \rightarrow N_1(\phi, 00) \xrightarrow{1} N_2(1, 0) \xrightarrow{0} N_3(10, \phi) \xrightarrow{1}$$

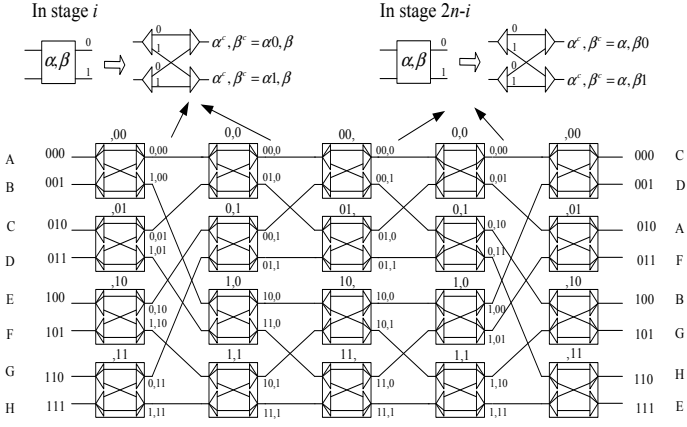

Fig. 5. Numbering of decomposed Benes network

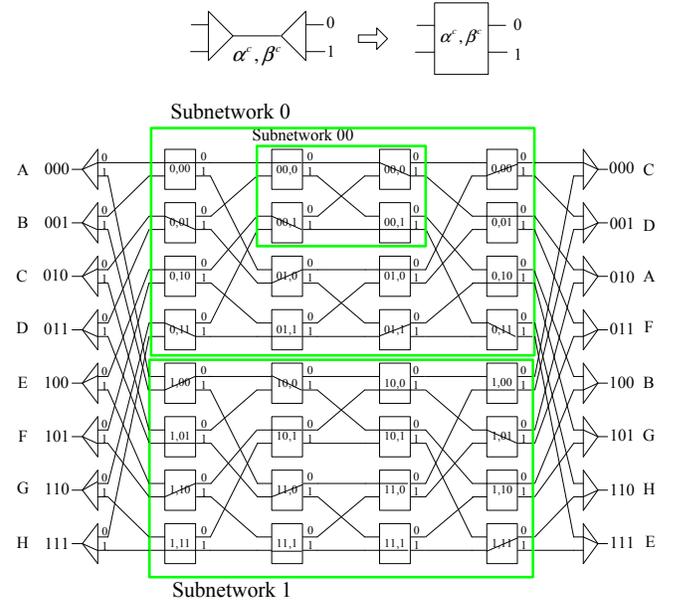

Fig. 6. Numbering of conjugate Benes network

$$N_4(1,1) \xrightarrow{0} N_5(\phi, 10) \xrightarrow{0} D(100),$$

with link sequence 10100.

### B. Conjugate Benes networks

The node numbering of conjugate Benes network is inherent from that of Benes network. The conjugate transformation converts an internal link of the Benes network to a node in the conjugate Benes network, in which the node numbering is the composite of its original link number and the attached node number. As shown in Fig. 5, the output link $a_i$ of node $N_i(a_1 \ldots a_{i-1}, \ b_1 \ldots b_{n-i})$ for $i = 1, \ldots, n-1$ in the baseline network, is labelled by $(a_1 \ldots a_i, \ b_1 \ldots b_{n-i})$. Similarly, the output link $b_{n-i+1}$ of node $N_{2n-i}(a_1 \ldots a_{i-1}, b_1 \ldots b_{n-i})$ for $i = 2, \ldots, n$ in the reverse baseline network is labelled by $(a_1 \ldots a_{i-1}, \ b_1 \ldots b_{n-i}b_{n-i+1})$. There is a one-to-one correspondence between the the output link of the node $N_k(\alpha, \beta)$ for $k = 1, \ldots, 2n-2$ in the Benes networks and the node of conjugate Benes network, as shown in Fig. 6. The conversion of the output link label of the node $N_k(\alpha, \beta)$ to the numbering $M_k(\alpha^c, \beta^c)$ of a merged node is governed by the following rule.

i) The output link 0 of the node $N_k(\alpha, \beta)$ is converted to

$$M_k(\alpha^c, \beta^c) = \begin{cases} M_k(\alpha 0, \beta) & \text{for } k = 1, \ldots, n-1, \\ M_k(\alpha, \beta 0) & \text{for } k = n, \ldots, 2n-2; \end{cases}$$

ii) the output link 1 of the node $N_k(\alpha, \beta)$ is converted to

$$M_k(\alpha^c, \beta^c) = \begin{cases} M_k(\alpha 1, \beta) & \text{for } k = 1, \ldots, n-1, \\ M_k(\alpha, \beta 1) & \text{for } k = n, \ldots, 2n-2. \end{cases}$$

Each node of the conjugate Benes network shown in Fig. 6 is also labelled by a 2-tuple of $n$ bits total from top to bottom, in which the first and second binary number are still the top-down numberings of the subnetwork and the node within the subnetwork, respectively. However, the smallest subnetwork is no longer a central module but a $4 \times 4$ two-stage network, called the *central subnetwork*. Thus, a path passing through the central module $(x_1 \ldots x_{n-1}, \phi)$ in the Benes network will become one passing through the central subnetwork $(x_1 \ldots x_{n-1})$ in the conjugate Benes network. For example, the path of connection B in the conjugate Benes

network is given as follows:

$$S(001) \xrightarrow{1} M_1(1,00) \xrightarrow{0} M_2(10,0) \xrightarrow{1} M_3(10,1) \xrightarrow{0} M_4(1,10) \xrightarrow{0} D(100),$$

in which nodes $M_2(10,0)$ and $M_3(10,1)$ belong to the central subnetwork 10. In general, the symbolic expression of the path from the source $S(s_1 \ldots s_n)$ to the destination $D(d_1 \ldots d_n)$ and passing through the selected central subnetwork $(x_1 \ldots x_{n-1})$ is given by:

i) The first sub-path:

$$S(s_1 \ldots s_n) \xrightarrow{x_1} \underbrace{M_1(x_1, s_1 \ldots s_{n-1})}_{\text{stage } 1} \xrightarrow{x_2}$$
$$\ldots \xrightarrow{x_i} \underbrace{M_i(x_1 \ldots x_i, s_1 \ldots s_{n-i})}_{\text{stage } i} \xrightarrow{x_{i+1}} \ldots \xrightarrow{x_{n-1}}$$
$$\underbrace{M_{n-1}(x_1 \ldots x_{n-1}, s_1)}_{\text{stage } n-1};$$

ii) the second sub-path:

$$\underbrace{M_{n-1}(x_1 \ldots x_{n-1}, s_1)}_{\text{stage } n-1} \xrightarrow{d_1} \underbrace{M_n(x_1 \ldots x_{n-1}, d_1)}_{\text{stage } n} \xrightarrow{d_2}$$
$$\ldots \xrightarrow{d_{n-i+1}} \underbrace{M_{2n-i}(x_1 \ldots x_{i-1}, d_1 \ldots d_{n-i+1})}_{\text{stage } 2n-i} \xrightarrow{d_{n-i+2}}$$
$$\ldots \xrightarrow{d_{n-1}} \underbrace{M_{2n-2}(x_1, d_1 \ldots d_{n-1})}_{\text{stage } 2n-2} \xrightarrow{d_n} D(d_1 \ldots d_n).$$

in which $M_{n-1}(x_1 \ldots x_{n-1}, s_1)$ and $M_n(x_1 \ldots x_{n-1}, d_1)$ are nodes of the central subnetwork $(x_1 \ldots x_{n-1})$, and the link sequence $x_1 \ldots x_{n-1}d_1 \ldots d_n$ remains the same.

In respect to optical switching, the most important property of conjugate Benes networks is the crosstalk-free route assignments given in the following theorem.

**Theorem 1:** The nonblocking route assignments of a Benes network become crosstalk-free in the conjugate Benes network.

*Proof:* The theorem is proved by contradiction. Under the conjugate transformation, suppose two nonblocking connection paths $X$ and $X'$ in a Benes network cross at the same node in the conjugate Benes network. Let $X$ and $X'$ be paths from input $S(s_1 \ldots s_n)$ to output $D(d_1 \ldots d_n)$, and input $S'(s'_1 \ldots s'_n)$ to output $D'(d'_1 \ldots d'_n)$, passing through central modules $(x_1 \ldots x_{n-1}, \phi)$ and $(x'_1 \ldots x'_{n-1}, \phi)$, respectively, in a Benes network. According to the numbering scheme of the conjugate Benes network, the following should hold if they cross at the same node at stage $i$ for $i = 1, \ldots,$ or $n-1$:

$$M_i(x_1 \ldots x_{i-1}x_i, s_1 \ldots s_{n-i})$$
$$= M_i(x'_1 \ldots x'_{i-1}x'_i, s'_1 \ldots s'_{n-i}), \qquad (1)$$

which implies the following identities should be held simultaneously in the Benes network:

$$N_i(x_1 \ldots x_{i-1}, s_1 \ldots s_{n-i})$$
$$= N_i(x'_1 \ldots x'_{i-1}, s'_1 \ldots s'_{n-i}) \qquad (2)$$

and

$$x_i = x'_i. \qquad (3)$$

That is, the two paths $X$ and $X'$ will penetrate the same node $N_i(x_1 \ldots x_{i-1}, s_1 \ldots s_{n-i}) = N_i(x'_1 \ldots x'_{i-1}, s'_1 \ldots s'_{n-i})$ at stage $i$, and pass through the same output link $x_i = x'_i$ in the Benes network, a contradiction with the nonblocking assumption on these two paths. Same result can be obtained if the two paths cross at the same node at stage $2n - i$ for $i = 2, \ldots,$ or $n$ in the conjugate Benes network. Thus, the theorem is established. $\qquad \square$

The above theorem reveals the fact that these nonblocking route assignments and rearrangements algorithms can be extended to crosstalk-free optical switches in a straightforward manner. The application to the three-stage Clos network is discussed in the next section, and the furthermore generalization to optical multicast switches is addressed in the section V.

## IV. THE CONJUGATE CLOS NETWORK

The conjugate transformation can be applied to any connecting networks. The $N \times N$ Clos network shown in Fig. 7 with $k$ input/output modules and $m$ central modules satisfies the rearrangeable nonblocking condition if $m \geq n$ [31], where $n$ is the number of input ports or output ports of each input module or output module, respectively. The construction of crosstalk-free Clos network is described below to illustrate the generalized cojugate transformation.

The nodes and links of the Clos network depicted in Fig. 7 are numbered as usual. According to this numbering scheme, the source $S$ is the input link $s_2$ of the input module $s_1$, in which $s_1 = \lfloor \frac{S}{n} \rfloor$ and $s_2 = [S]_n$, where $\lfloor \frac{a}{b} \rfloor$ is the largest integer smaller than $a/b$ and $[a]_b$ is the remainder of $a/b$. Similarly, the destination $D$ is the output link $d_2$ of the output module $d_1$, where $d_1 = \lfloor \frac{D}{n} \rfloor$ and $d_2 = [D]_n$. A connection passes through the central module $x_1$ from the source $S(s_1, s_2)$ to the destination $D(d_1, d_2)$ can be expressed as follows:

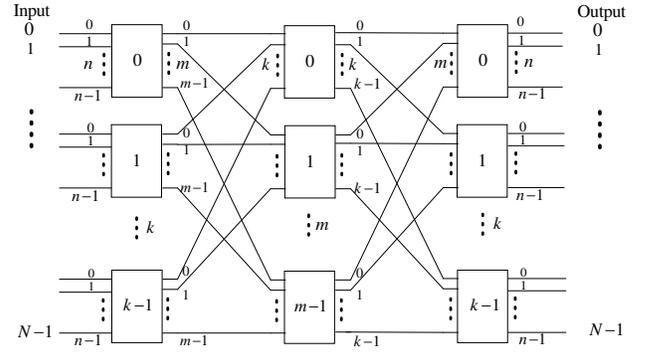

Fig. 7. Numbering of three-stage Clos network

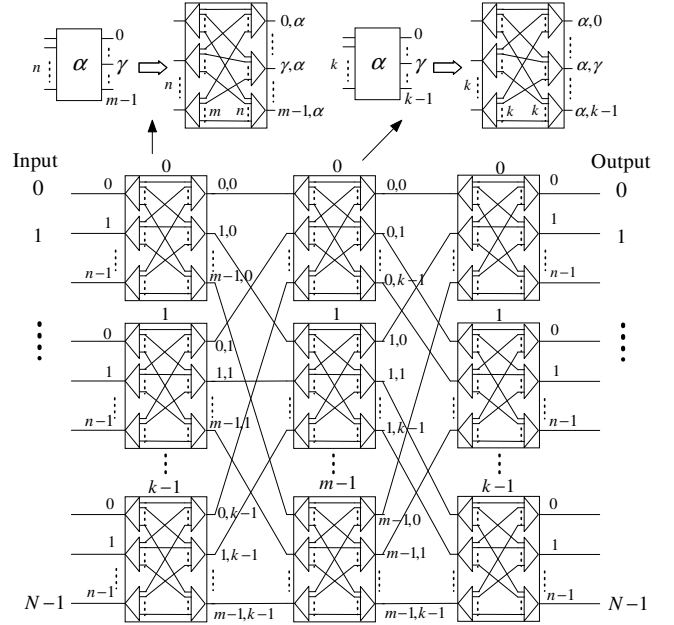

Fig. 8. Numbering of decomposed Clos network

$$S(s_1, s_2) \longrightarrow \underbrace{s_1}_{\text{stage 1}} \xrightarrow{x_1} \underbrace{x_1}_{\text{stage 2}} \xrightarrow{d_1} \underbrace{d_1}_{\text{stage 3}} \xrightarrow{d_2} D(d_1, d_2).$$

In the decomposed Clos network shown in Fig. 8, it is clear that the output link $\gamma$ of a first or second stage node $\alpha$ can be labelled by $(\gamma, \alpha)$ or $(\alpha, \gamma)$, respectively, according to our numbering rule. Consequently, as shown in Fig. 9, each merged node in the conjugate Clos network will naturally adopt the corresponding link label of the original Clos network. The numbers in the 2-tuple label of each node represent the subnetwork and the node within the subnetwork, respectively. In this conjugate network, the previous connection passing through the subnetwork $x_1$ is expressed by:

$$S(s_1, s_2) \xrightarrow{x_1} \underbrace{(x_1, s_1)}_{\text{stage 1}} \xrightarrow{d_1} \underbrace{(x_1, d_1)}_{\text{stage 2}} \xrightarrow{d_2} D(d_1, d_2).$$

In the following, we show that the conjugate Clos network possesses crosstalk-free property under the above transformation.

*Theorem 2:* The nonblocking route assignments of a three-stage Clos network become crosstalk-free in the conjugate Clos network.

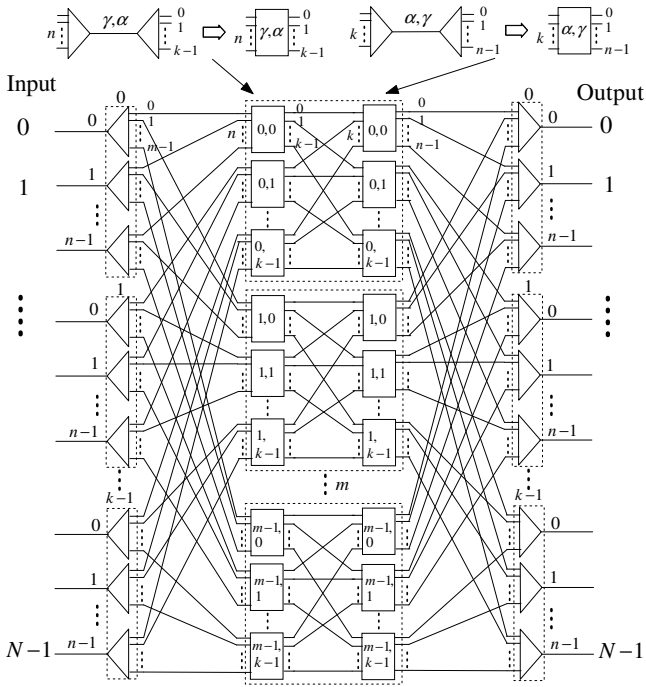

Fig. 9. Numbering of conjugate Clos network

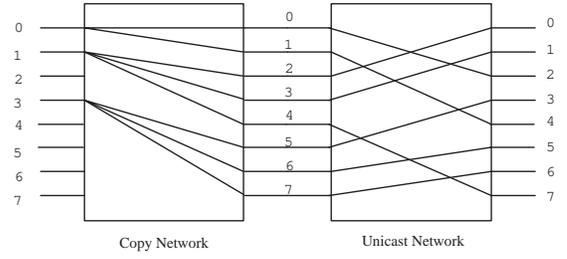

Fig. 10. Multicasting by copy network

*Proof:* Suppose two nonblocking $X$ and $X'$ in the original Clos network cross at the same node in the conjugate Clos network. Let $X$ and $X'$ be paths from input $S(s_1, s_2)$ to output $D(d_1, d_2)$, and input $S'(s_1', s_2')$ to output $D'(d_1', d_2')$, passing through central modules $x_1$ and $x_1'$, respectively, in the Clos network. Under the conjugate transformation, the following should hold if they cross at the same node in the first stage of the conjugate Clos network:

$$x_1', s_1' = x_1', s_1', \qquad (4)$$

which implies that they pass through the same link $x_1 = x_1'$ on the same node $s_1 = s_1'$ in the first stage of the Clos network, a contradiction with the nonblocking assumption. Similarly, it is impossible that the two paths will cross at the same node in the second stage of the conjugate Clos network. □

Since $m = n$ is the minimum requirement for rearrangeable crosstalk-free routing, it is easy to show that the total number of switch elements in the cojugate Clos network is equal to $2nk + N = 3N$. Apply the same decomposition to the general $(2\log_d N - 1)$-stage Benes network, constructed by $d \times d$ switch elements, resulting $(2\log_d N - 1)N$ nodes in its conjugate network.

## V. Crosstalk-free multicast switching networks

A multicast switch is usually realized by the cascaded combination of two networks, as shown in Fig. 10, a copy network and a point-to-point switch network. The copy network replicates input signals and the point-to-point switch routes those resulting signals to their respective outputs. For example, the following set of multicast connection requests is realized by the network shown in Fig. 10 in two steps:

$$\begin{pmatrix} \text{Input} \\ \text{Output} \end{pmatrix} = \begin{pmatrix} 0 & 1 & 3 \\ (2,4) & (0,1,7) & (3,5,6) \end{pmatrix}, \qquad (5)$$

In the first step, copies of each input are generated and assigned to the following range addressed outputs monotonically to satisfy the nonblocking condition:

$$\begin{pmatrix} \text{Input} \\ \text{Output} \end{pmatrix} = \begin{pmatrix} 0 & 1 & 3 \\ (0,1) & (2,3,4) & (5,6,7) \end{pmatrix}. \qquad (6)$$

In the next step, the point-to-point switch will establish the following connections:

$$\begin{pmatrix} \text{Input} \\ \text{Output} \end{pmatrix} = \begin{pmatrix} 0 & 1 & 2 & 3 & 4 & 5 & 6 & 7 \\ 2 & 4 & 0 & 1 & 7 & 3 & 5 & 6 \end{pmatrix}. \qquad (7)$$

Again, a realization of muticast switch based on Benes network will be discussed next to illustrating the generalized nonblocking copy process based on rank-based assignment algorithm and interval splitting algorithm proposed in [25], [26].

### A. Benes copy networks

A three-stage Clos network is rearrangeable nonblocking if its number of central modules is greater than or equal to the port number of its input/output modules. The route assignment problem is equivalent to edge-coloring of a bipartite graph, in which two disjoint sets of nodes represent the input and output modules, respectively, and each edge represent a connection between an input and an output. As shown in Fig. 11, the central module assigned to a connection corresponds to the color on each edge. It has shown in [25], [26] that if the set of connection requests is monotonic, the corresponding bipartite graph can be edge-colored by a rank-based algorithm. In the example shown in Fig. 11, the set of connection requests A∼H is monotonic, and the corresponding edges are colored by I, II, III, IV, I, II, III, and IV, respectively.

The rank-based algorithm can be iteratively applied to Benes network, because it is recursively constructed from Clos network. In this section, an extension of the Benes network, called *Benes copy network*, is proposed in conjunction with the generalization of rank-based algorithm to implement monotonic connections requesting for plural outputs.

Let $S_0, S_1, \ldots, S_{k-1}$, where $S_m > S_n$ if $m > n$, be a set of active inputs, and $R_0, R_1, \ldots, R_{k-1}$ be the corresponding is concentrated by the baseline network, while replications of signals are carried out by the reverse baseline network. Thus, each multicast connection consists of a path from the input to the assigned central module and a binary tree from the central module to outputs.

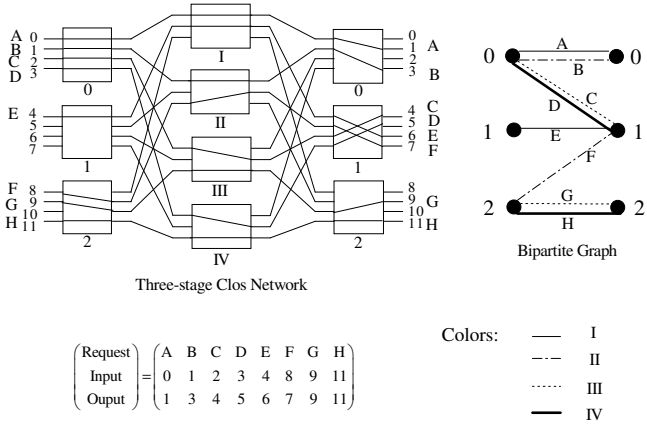

Fig. 11. Rank-based Assignments Algorithm in three-stage Clos network

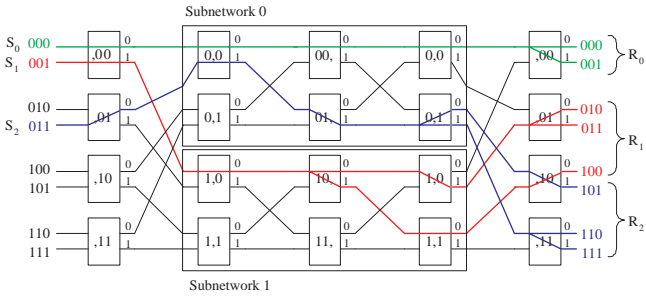

Fig. 12. Benes copy network

**TABLE I**
**Table of routing tags**

| **S** | **0** | **1** | **3** |
|---|---|---|---|
| **R** | **0,1** | **2,3,4** | **5,6,7** |
| **rank** | **0** | **1** | **2** |
| $a_3a_2a_1$ | **000** | **001** | **010** |
| Interval (min, max) | **(000,001)** | **(010,100)** | **(101,111)** |
| **Routing tags** $=a_1a_2$ | **00** | **10** | **01** |
| Interval (min, max) | **(000,001)** | **(010,100)** | **(101,111)** |

The routing in the baseline network is determined by the labels of central modules assigned to input call requests. The destination addresses of each input signal is an interval, specified by a pair of binary numbers $minimum$ and $maximum$, which is determined by the running-sum of copies requested. The *Interval Splitting Algorithm* is performed in reverse baseline network to replicate signals and route them to the outputs within the range of address interval. A description of the interval splitting algorithm is given in appendix II. Table I lists the ranks and address intervals of the set of connection requests given in (6), and they serve as the respective routing tags, explained above, of these connections in baseline network and reverse baseline networks.

### B. Conjugate Benes Copy networks

In respect to crosstalk problems, the decomposition and merge procedures can be effectively applied to copy networks as well. As shown in Fig. 13, the three states of each $2 \times 2$ node of a copy network remain the same under the transformation. In bar and cross states, crosstalk signals are carried by separate nodes after the transformation, and the broadcast signals in copy state are not concerned in crosstalk anyway. An $8 \times 8$ conjugate Benes copy network resulting from the decomposition and merge procedures is depicted in Fig. 14. Again, it is shown in the following theorem that the rank-based route assignments are crosstalk-free in the conjugate Benes network.

*Theorem 3:* The nonblocking route assignments of a Benes copy network become crosstalk-free in the conjugate Benes copy network.

*Proof:* Suppose the two connection requests $X : (S_i, R_i)$ and $X' : (S_j, R_j)$ are ranked $i = a_n \ldots a_1$ and $j = b_n \ldots b_1$ in a monotonic sequence of requests, and therefore being assigned with central modules $(a_1 \ldots a_{n-1}, \phi)$ and $(b_1 \ldots b_{n-1}, \phi)$, respectively. Since the rank-based assignments are nonblocking, the two paths $(S_i, r_i)$ and $(S_j, r_j)$ should be link-disjoint in Benes network. According to Theorem 1, the paths of $(S_i, r_i)$ and $(S_j, r_j)$ are node-disjoint after conjugate transformation, for any $r_i \in R_i$ and $r_j \in R_j$. It follows that the two mulicast connections $(S_i, R_i)$

set of plural outputs. This set of multicast call requests is monotonic if

$$S_m < S_n \;\Rightarrow\; r_m < r_n \quad \forall r_m \in R_m \text{ and } \forall r_n \in R_n \quad \text{or;}$$
$$S_m > S_n \;\Rightarrow\; r_m > r_n \quad \forall r_m \in R_m \text{ and } \forall r_n \in R_n.$$

The basic function of a copy network is to generate exact number of copies requested by each input, the switching function will be carried out by the point-to-point switches in the subsequent stage. Without loss of generality, we assume that each set of outputs $R_m$, $m = 0, ..., k-1$ is a consecutive address interval.

The rank of a call request $(S_i, R_i)$ is its index $i$ in the monotonic sequence $\{(S_0, R_0), \ldots, (S_{k-1}, R_{k-1})\}$. Let $i = a_n a_{n-1} \ldots a_2 a_1$, where $n = log_2 N$, be the binary number representation of the rank $i$, the *Rank-based Assignment Algorithm of Benes copy networks* will assign the central module labelled by $(a_1 a_2 \ldots a_{n-1}, \phi)$ to the call request $(S_i, R_i)$, for $i = 0, \ldots, k-1$. This assignment is nonblocking, and the proof is provided in the appendix I. As an example, the binary number representation of ranks of call requests $(S_0, R_0)$, $(S_1, R_1)$ and $(S_2, R_2)$ are $0 = 000, 1 = 001$, and $2 = 010$, respectively, that correspond to the central module $(00, \phi)$, $(10, \phi)$ and $(01, \phi)$ assigned to them as shown in Fig. 12.

As we mentioned before, the Benes network is a combination of a baseline network, from inputs to the middle stage, and the reverse baseline network, from the middle stage to outputs. In the Benes copy network, the set of call requests

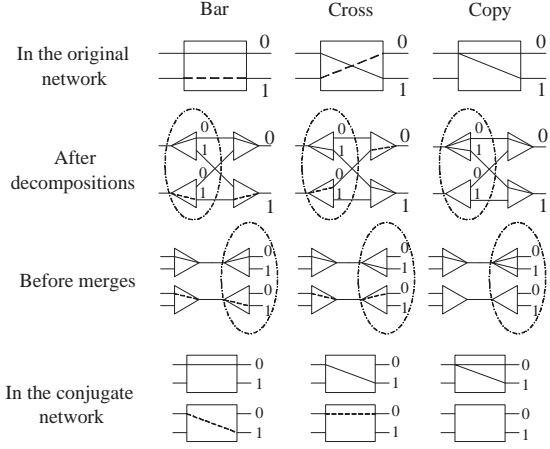

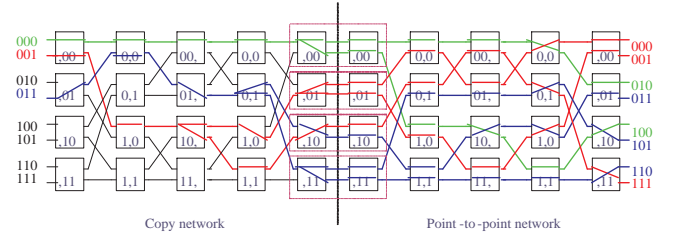

(a) Network I

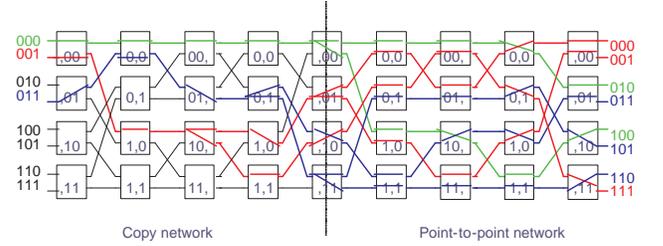

(b) Network II

Fig. 13. Conjugate transformation of Benes copy network

Fig. 15. Multicast Benes network

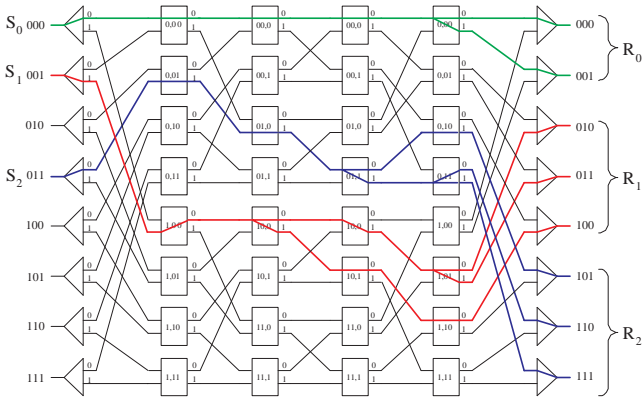

Fig. 14. Conjugate Benes copy network

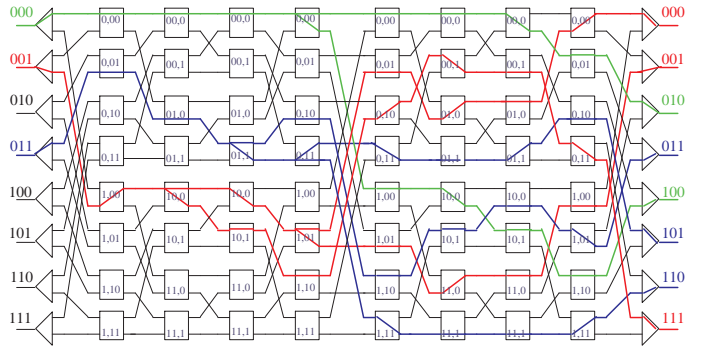

Fig. 16. Conjugate Multicast Benes network

and $(S_j, R_j)$ are crosstalk-free in the conjugate Benes copy network.

□

### C. Crosstalk-free multicast switching networks

A multicast switch can be synthesized by the cascaded combination of two Benes networks. As an example, the set of multicast connection requests given in (5) is established in the multicast Benes network shown in Fig. 15(a). The last stage of Benes copy network and the first stage of Benes network can be combined into one stage to reduce the redundancy as shown in Fig. 15(b), resulting in a $(4n-3)$-stage multicasting network. It is assured by theorem 1 and 2, the nonblocking route assignments in the multicast Benes network become crosstalk free in the conjugate multicasting network shown in Fig. 16.

## VI. CONCLUSION

DC-based high speed scalable photonic switches suffer from crosstalk problems when two optical signals cross at the same DC element. In the past serval decades, the nonblocking electronic switching networks have been widely studied and a mature theory has been erected. An easy-to-implement conjugate transformation that turns nonblocking route assignments into crosstalk-free in the corresponding conjugate network is described in this paper, and we present a generic and systematic approach to design and implement crosstalk-free switches that is parallel to the nonblocking switching theory. We also show that this crosstalk-free design principle can be further extended to multicast switches in a straightforward manner.

## APPENDIX I
## PROOF OF RANK-BASED ASSIGNMENT ALGORITHM OF BENES COPY NETWORKS

*Proof:* Suppose two replication requests, $(S_i, R_i)$ input from $S_i(s_1 \ldots s_n)$ and $(S_j, R_j)$ input from $S_j(s'_1 \ldots s'_n)$, ranked by $i = a_n \ldots a_1$ and $j = a'_n \ldots a'_1$, select central modules $(a_1 \ldots a_{n-1}, \phi)$ and $(a'_1 \ldots a'_{n-1}, \phi)$, respectively.

If they collide at an output link of a node at stage $k$ for $k = 1, \ldots, n-1$ in the baseline network, then according to the numbering of Benes networks we have:

$$N_i(a_1 \ldots a_{k-1}, s_1 \ldots s_{n-k}) = N_i(a'_1 \ldots a'_{k-1}, s'_1 \ldots s'_{n-k}) \tag{8}$$

and

$$a_k = a'_k, \tag{9}$$

which lead to the following identities:

$$a_1 \ldots a_k = a'_1 \ldots a'_k \tag{10}$$

and

$$s_1 \ldots s_{n-k} = s'_1 \ldots s'_{n-k}. \tag{11}$$

Since $i$ and $j$ are the ranks of $(S_i, R_i)$ and $(S_j, R_j)$, respectively, in a monotonic sequence of connection requests, we have

$$|j - i| \leq |S_j - S_i|. \tag{12}$$

From (10) and (11), we obtain

$$
\begin{aligned}
|j - i| &= |a'_n \ldots a'_1 - a_n \ldots a_1| \\
&= 2^k |a'_n \ldots a'_{k+1} - a_n \ldots a_{k+1}| \\
&\geq 2^k,
\end{aligned}
\tag{13}
$$

and

$$
\begin{aligned}
|S_j - S_i| &= |s'_1 \ldots s'_n - s_1 \ldots s_n| \\
&= |s'_{n-k+1} \ldots s'_n - s_{n-k+1} \ldots s_n| \\
&\leq 2^k - 1.
\end{aligned}
\tag{14}
$$

But (12), (13) and (14) together imply $0 \leq -1$, a contradiction.

Similarly, it is impossible that any two paths $(S_i, r_i)$ and $(S_j, r_j)$ would collide at an output link of a node in the reverse baseline network, for $r_i \in R_i$ and $r_j \in R_j$. It follows that the connection trees generated by requests $(S_i, R_i)$ and $(S_j, R_j)$ are link-disjoint in both baseline and reverse baseline networks, and the set of rank-based assignments is nonblocking in the Benes copy network.

□

## Appendix II
## Interval splitting algorithm

Suppose each input requests for plural outputs within the range of an address interval, which is delineated by two numbers, *minimum* and *maximum*. The replications are carried out in the reverse baseline network from the middle stage $n$ to the last stage $2n - 1$. Initially, the address interval is represented by $\min(n - 1) = minimum$ and $\max(n - 1) = maximum$, which will be modified in each subsequent stage during the course of replication process. At stage $i$ for $i = n, \ldots, 2n - 1$, the address interval $\min(i - 1) = v_n \ldots v_{2n-1}$ and $\max(i - 1) = V_n \ldots V_{2n-1}$ provides the instruction for the node to conduct the following operations:

i) If $v_i = V_i = 0$ or $v_i = V_i = 1$, then send the request out on link 0 or 1, respectively.

ii) If $v_i = 0$ and $V_i = 1$, then send out the request on both links with the following updated address intervals.

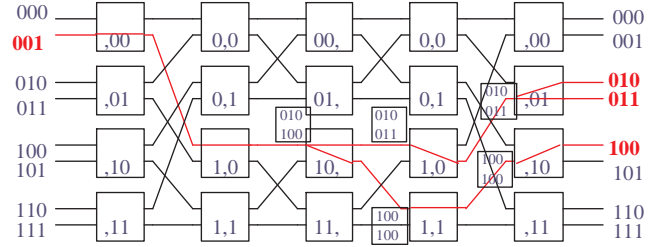

Fig. 17. Replication process in Benes copy network

For the request sent out on link 0,

$$
\begin{aligned}
\min(i) &= \min(i - 1) \\
&= (v_n \ldots v_{2n-1}) \\
\max(i) &= (v_n \ldots v_{i-1}01 \ldots 1).
\end{aligned}
\tag{15}
$$

For the request sent out on link 1,

$$
\begin{aligned}
\min(i) &= (V_n \ldots V_{i-1}10 \ldots 0) \\
\max(i) &= \max(i - 1) \\
&= (V_n \ldots V_{2n-1}).
\end{aligned}
\tag{16}
$$

The above updating routine splits the address interval into two sub-intervals, each of them specifies output range of a sub-tree of the original tree. It should be noted that the above set of rules implies that $v_j = V_j$, for $j = n, \ldots, i-1$, in the address interval, and the event $v_i = 1$ and $V_i = 0$ is impossible due to the min-max representation of addresses intervals at stage $i$. An example to illustrate this Interval Splitting Algorithm is provided in Fig. 17.